\pdfoutput=1

\documentclass[aps,pra,preprint,groupedaddress]{revtex4}

\usepackage{graphicx}
\usepackage{verbatim}

\begin{document}


\title{An Analytical Error Model for Quantum Computer Simulation}


\author{Eric Chi}
\author{Stephen A. Lyon}
\author{Margaret Martonosi}
\affiliation{Dept. of Electrical Engineering, Princeton University}
\email[]{(echi,lyon,mrm)@princeton.edu}


\date{\today}

\begin{abstract}
Quantum computers (QCs) must implement quantum error correcting codes (QECCs)
to protect their logical qubits from errors,
and modeling the effectiveness of QECCs on QCs is an important problem
for evaluating the QC architecture.
The previously developed Monte Carlo (MC) error models
may take days or weeks of execution to produce an accurate result
due to their random sampling approach.
We present an alternative analytical error model
that generates, over the course of executing the quantum program, 
a probability tree of the QC's 
error states.
By calculating the fidelity of the quantum program directly,
this error model has the potential for enormous speedups over the MC model
when applied to small yet useful problem sizes.
We observe a speedup on the order of 1,000X when accuracy is required,
and we evaluate the scaling properties of this new analytical error model.

\end{abstract}


\maketitle

\section{Introduction}
\label{sec:intro}

Errors from decoherence and gate imprecision are one of the greatest 
challenges in realizing a quantum computer.
Fortunately, quantum error correction and fault-tolerant protocols have been developed to combat decoherence in 
quantum computers (QCs)~\cite{Preskill:fault-tolerant,Shor:FT,Steane:qecc,Steane:active_stabilisation,Steane:fault-tolerant}.
These approaches encode {\em logical} program qubits into code blocks and perform all logical operations on the encoded blocks so as to detect and correct errors and prevent any errors from propagating wildly.
For example, the [[7,1,3]] quantum error correction code (QECC)
encodes one logical qubit into a block of seven {\em physical} qubits.
Every logical operation is followed by an error recovery phase
that extracts error syndromes from a logical qubit code block
and performs corrective procedures if necessary.

Error modeling is critical for developing and evaluating QC architectures.
By tracking the probabilities of errors acting on physical qubits,
we may evaluate the effectiveness of  
QECCs, error recovery techniques, and microarchitecture noise tolerances.

The simplest noise models apply the Monte Carlo (MC) simulation strategy, which
randomly samples possible error scenarios using a random number generator.
This methodology has been applied to previous studies 
including~\cite{Balensiefer:framework,Steane:overhead}.
The downside to MC simulation is its long runtime. 
When used to measure the probability of an uncommon event 
(for example, the probability that error recovery will fail),
the MC simulator will need to run a number of iterations 
that is several orders of magnitude more than the expected period of that event
in order to generate enough samples.
For example, a small quantum program with only two logical qubits
required nearly eight days of MC simulation runtime 
to achieve three significant digits of accuracy.

We present an alternative error model that analytically tracks
error probabilities rather than relying on random sampling.
Since random sampling potentially requires many tries before
encountering rare events,
our error model aims to improve performance by
analytically computing probabilities.
Our model tracks the possible error states of the simulated QC
and mitigates the exponential growth in the number of error states
by pruning the probability tree with customizable thresholds.
We compare the effectiveness of this analytical error model 
with a simple MC model; 
focus on speed versus accuracy trade-offs.


Steane's work~\cite{Steane:overhead} presented a similar approach 
with a probability tree-based error model verified with MC simulation.
Our work distinguishes itself by being more general and fitting into a simulation framework.
Whereas Steane's model is a formula to calculate the crash rate probability based on the structure of the QECC check matrices,
our error model processes generic error events and may be
better suited for analyzing different error recovery protocols
or architectural overheads.

This paper proceeds as follows:
Section~\ref{sec:comb} describes our analytical error model;
Section~\ref{sec:method} describes our experimental methodology;
Section~\ref{sec:results} presents our results;
and Section~\ref{sec:conclusion} concludes.

\section{\label{sec:comb}An analytical approach to quantum error simulation}

\subsection{Algorithm overview}

The overall algorithm of our analytical error model
is to develop a probability tree that tracks the
evolution of possible errors in the 
physical qubits comprising the quantum computer. 
By tracking the emergence and evolution of random errors,
the model may calculate the effectiveness of quantum error correction
and the fidelity of the quantum program.

Each node in this probability tree pairs an error state,
which represents a possible error condition for a set of qubits, 
to a probability value. 
The probability tree is initialized to a known starting state.
From the error model's perspective, 
a quantum program is represented as a series of 
error {\em events} and error {\em tasks}.
These error events or tasks evolve the probability tree
by creating new leaf nodes 
(and potentially expanding the number of possible error states) 
reflecting the desired state and probability changes.
The final probability tree will have a number of levels
equal to the number of error events and tasks processed.
In practice, the implementation need only track the most recent level of
the probability tree to generate the next level.

Upon conclusion of program simulation, the error model sums the probabilities
of the leaf nodes that match desired characteristics
(e.g., where the code blocks sustain no more than a correctable number 
of errors).
The error model may thereby calculate the overall failure rate of the program.

\subsection{Pauli strings as error states}

An arbitrary error state of an individual qubit 
may be quantized into one of the four Pauli operators:
$I$, $X$, $Y$, or $Z$. 
The identity operator $I$ indicates the error-free state. 
The $X$, $Z$, and $Y$ error states
indicate the presence of a bit-flip, phase-flip, and combined bit- and phase-flip errors, respectively. 
Each error state is associated with a probability value,
and a qubit may be in a superposition of multiple error states.

It is necessary to track joint error probabilities amongst qubits
to correctly calculate correlated errors.
Our model accomplishes this by tracking the error states of a set
of qubits as a group encapsulated in a {\em QubitSet} data structure.
The error states in a QubitSet are then represented as 
strings of Pauli operators (or Pauli strings).
In our error model, each physical qubit is associated with
exactly one QubitSet,
so the QubitSets partition all of the qubits in the simulated QC.

In our implementation, every QubitSet has a hash table that associates the 
Pauli string error states to double floating-point probability values:
an {\em error map}.
The states in this error map represent the potential error conditions
for the qubits in the QubitSet.
The error map effectively contains all of the nodes on the same level of the
error probability tree.
The transition from one level to the subsequent level of the probability tree
is modeled by error events and error tasks that evolve error maps.

\subsection{Error events and tasks represent the quantum program}

Error events and error tasks manipulate the error maps to reflect 
potential changes in state.
Error events represent potential sources of errors during execution.
The set of possible events is open-ended,
but we incorporated the following potential errors:
qubit decoherence during idle time (memory errors),
decoherence while a qubit is moving (transportation errors),
decoherence during operations,
and errors due to operation imprecision. 
Our error events affect either one or two qubits 
(to represent correlated errors from two-qubit operations).
If an error event targets two qubits, both qubits need to be members
of the same QubitSet.

Error tasks are more general actions that interact with error maps.
These tasks may modify error maps similarly to error events,
or they may simply iterate through the entries to make an interesting 
calculation.
Because error correction is such an important application for 
error analysis,
many of the error tasks that we have implemented
enact error correction effects.
For example, we have tasks that represent ancilla verification, 
syndrome extraction, and the actual correction of data code blocks.
These tasks are tied to measurement operations when information is revealed.

Another set of important tasks direct QubitSets to merge or split.
The number of potential states in a QubitSet's error map
is exponentially related to the number of qubits in that set.
Therefore, it is desirable to have smaller QubitSets when possible.
However, two-qubit error events require that both qubits exist
in the same QubitSet.
If the two target qubits of such an event reside in different QubitSets,
then those two QubitSets must merge into a larger QubitSet.
Splitting larger QubitSets into smaller QubitSets is desirable 
to reduce the overall memory footprint when the loss of 
joint probability information is acceptable.
This case often occurs after measurement of qubits.
The merging and splitting of QubitSets 
will be subsequently described in greater detail.

From the error model's perspective, 
a simulated quantum program is a sequence of these
error events and error tasks.
This error model can be embedded into an existing
QC architecture simulator by embedding these error events and tasks
into the program instructions, 
and much of this embedding may be automated.

\subsection{Evolving error maps}
Error events and many error tasks function by evolving a QubitSet's 
error map.
The basic algorithm is straightforward:
an empty error map is used as the new error map;
every state in the old error map is possibly transformed and/or expanded
(branched) into multiple states and then appended to the new error map.
This new error map encapsulates the next lower level of the error
probability tree.

We follow the example of~\cite{Steane:overhead} for the introduction
and propagation of errors.
An error event is stochastic; 
it leaves the QubitSet unchanged with probability $1-f$, the probability that the event does not trigger an error.
A one-qubit error event may transform an error state by adding an $X$, $Y$, or $Z$
error with equal probabilities summing to $f$.
Two-qubit error events fall into 
fifteen possibilities:
$IX$, $IY$, $IZ$, $XI$, $XX$, $XY$, $XZ$, $YI$, $YX$, $YY$, $YZ$, $ZI$, $ZX$, $ZY$, and $ZZ$. 
Certain operations may also transform or propagate errors.
For example, the Hadamard operation transforms $X$ to $Z$ errors and vice-versa.
Two-qubit operations such as the controlled-Not (CNot) gate may 
propagate certain errors between the two operand qubits.

\begin{figure}
\centering
\includegraphics[width=3.2in]{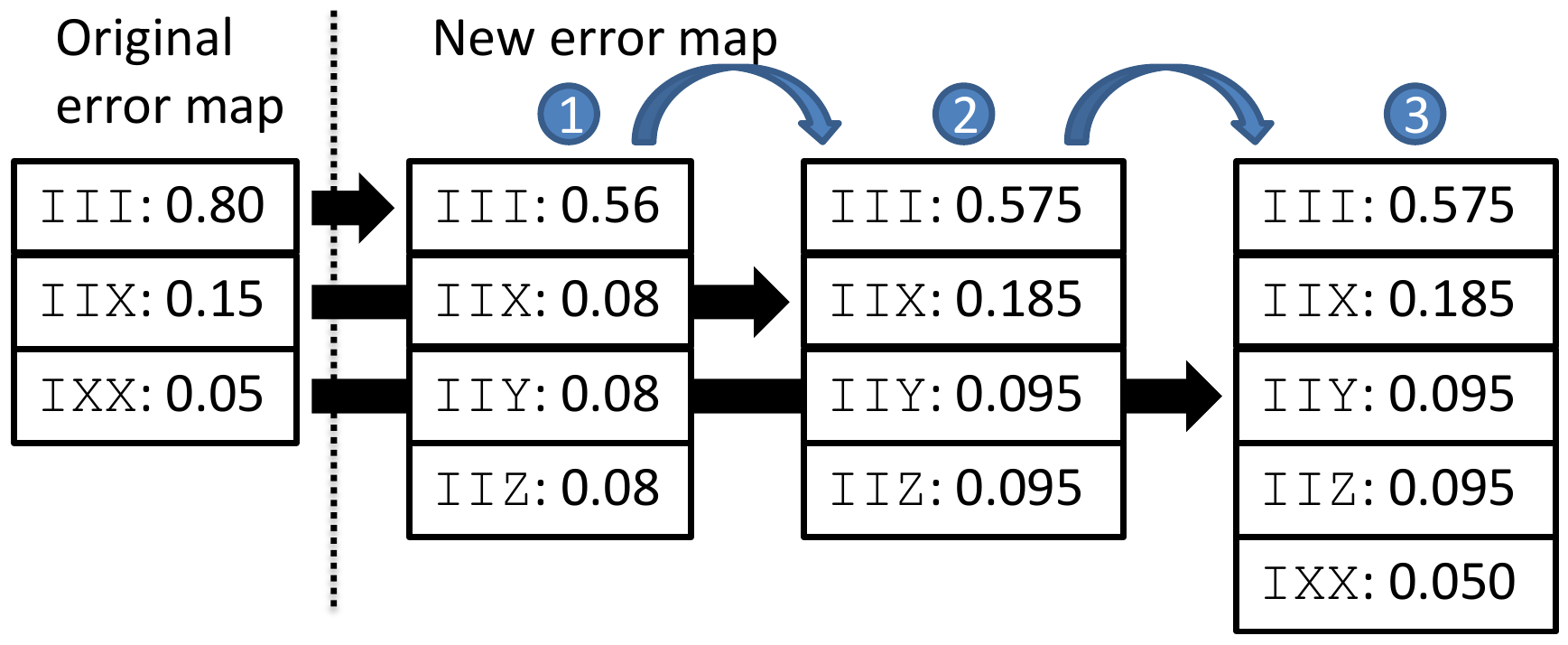}
\caption{
Example. 
An error event with probability 0.3
acts on bit 0 of a 3-bit QubitSet.
Assume that the event branch threshold is 0.1.
The noise event creates a new error map and 
(1) initially expands the error-free $III$ state into four states:
$III$, $IIX$, $IIY$, and $IIZ$.
(2) The original $IIX$ state is processed next and also expanded with its resultant error scenarios 
added to the new error map.
(3) The $IXX$ state falls below the event branch threshold so is appended
to the new error map without expansion.
}
\label{fig:noiseEvent}
\end{figure}

Because the branching state behavior during this evolution may lead to an
exponential increase in the number of states,
we impose a parameterized {\em event branching threshold}.
This branching threshold prevents states with probabilities smaller than the threshold from branching into even smaller states. 
An appropriately selected threshold will
abort the creation of minute error states that are relatively insignificant. 
The error map evolution process including the application of the branch threshold is illustrated in Figure~\ref{fig:noiseEvent}.

\subsection{\label{sec:comb:merging}Merging and splitting QubitSets}

Merging is a special task that takes two QubitSets as inputs and creates
a new combined QubitSet that contains all of the inputs' constituent qubits.
The merged QubitSet's error states are a cross product of the
input error maps' states.
A {\em merge threshold} is applied during this cross product
to reduce the state-space expansion,
so a new merged state is created only if its resultant probability is
greater than the merge branch threshold.

\begin{figure}
\centering
\includegraphics[width=2.5in]{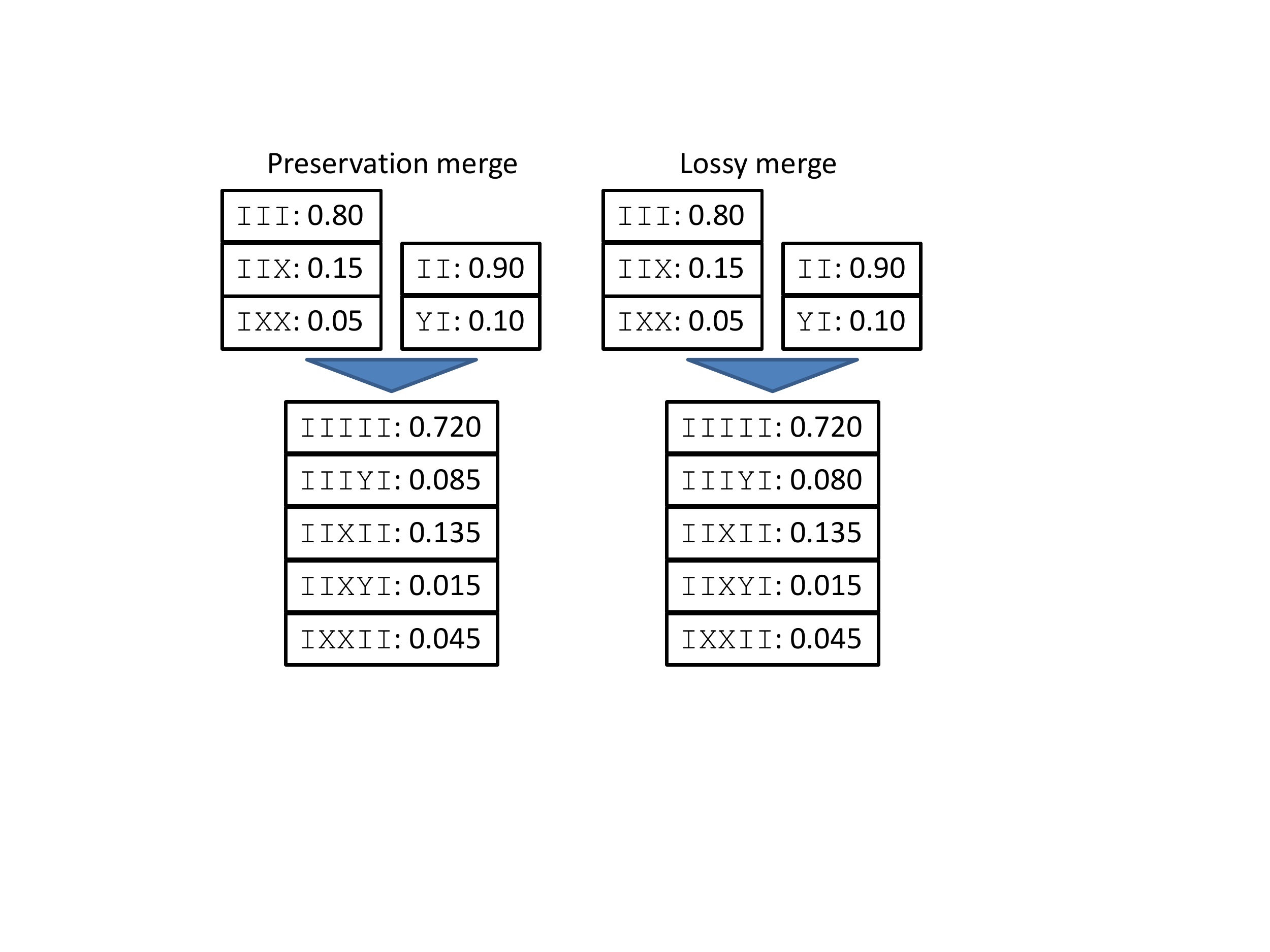}
\caption{
Two QubitSets merge with a merge threshold of 0.01.
The merged error state $IXXYI$ has a probability of 0.005, which is below the
merge threshold.
The preservation approach converts this merged state to $IIIYI$ and
avoids creating a new error state with minuscule probability.
On the other hand, the lossy approach simply discards this error state.
}
\label{fig:merge}
\end{figure}

We have two approaches to handling merged error states that fall below the
merge branch threshold:
the {\em preservation} approach and the {\em lossy} approach.
The preservation approach takes the less probable of the two states to be merged
and converts that state to the error-free state.
This way, the error information in the more probable error state is retained in the resultant merged state.
The lossy approach simply discards the low-probability merged error states.
These two approaches will yield near-similar results when the merge threshold
is set appropriately.
If the merge threshold is too large,
the preservation approach will overestimate the success rate 
while the lossy approach will underestimate the success rate.
This is because the less probable an error state is, the more errors it is 
likely to have (noise events are assumed to be improbable).
The preservation approach converts low probability states with 
higher error weights (the number of bits in a Pauli string bearing errors)
to lower error-weight states, reinforcing their probabilities.
These lower weight error states are more likely to be correctable by QECCs,
so the resultant success rate is inflated.
Conversely, the lossy approach discards error states and may lead to an
undercounting of successful error states.
Figure~\ref{fig:merge} illustrates merging QubitSets with these two approaches.

After qubits have been measured 
(in the course of simulated program execution), 
it is no longer necessary to track their error states.
A QubitSet may then {\em split} to remove the measured qubits and reduce
the size of its state space.
Splitting a QubitSet partitions a QubitSet into two smaller QubitSets
and is the inverse procedure of merging.
No thresholds are necessary for this procedure as the state space never
expands while splitting.

\subsection{Modeling error correction effects}

Because quantum error correction dominates the execution load of a QC,
the principal purpose of an error model is to evaluate the 
effectiveness of applying a QECC.
Our approach models the behavior of error correction by tracking the error
states of the encoded qubits and the ancilla qubits used in the error recovery
process.
If the error states of these qubits are within the tolerances of the
QECC, then the error recovery procedure is considered a success.

Error recovery involves the following steps:
preparing several blocks of ancilla qubits to a specific state;
verifying that the ancillae were prepared correctly by 
interacting them with verification bits;
interacting the prepared ancilla blocks with the encoded data qubits
to extract syndromes from the data block;
and performing corrective operations on the data block based on the syndrome
information.

Although our error model is not meant to track program state,
in the case of error correction, the program state can be rather pertinent.
The principal example here is the syndrome extraction process.
After interacting the ancilla block with the data block,
the ancilla is measured, and that measured value is the syndrome.
That syndrome is important state information that we want to keep,
so we have a syndrome measurement error task 
that stores the syndrome information
in the error state space corresponding to the measured ancilla qubits.
This dual use of the error state allows the syndrome information
to be stored in the error map.
Corrections to the data block may be made as a variation of the general
error map evolution algorithm:
prepare a new error map; iterate through the old error map; 
make a decision to correct the data based 
on the stored syndrome information;
and
append the resultant error state to the new error map.
The other components of the error recovery process are similarly effected
via error tasks that modify the error map based on 
error state information.

\section{Simulation methodology}
\label{sec:method}

We programmed both our analytical error model and a 
simple MC error model in Java.
For concreteness, we adopted the eSHe QC architecture~\cite{Chi:tailoring} for noise and timing parameters (Table~\ref{tab:parameters}), but the modeling approach is applicable to any architecture.

\begin{table}
\begin{center}
\begin{tabular}{|l|c|}
\hline
movement speed & 100 $\mu$m/$\mu$s \\
1-bit op time & 1 $\mu$s  \\
2-bit op time & 1 ms \\
\hline
memory decay constant & $1\times10^5$ s \\
operation decay constant & $5\times10^3$ s \\
transportation decay constant & $2.5\times10^4$ s \\
1-qubit op error rate & $1\times10^{-6}$ \\
2-qubit op error rate & $1\times10^{-4}$ \\
measurement error rate & $1\times10^{-4}$ \\
reset error rate & $1\times10^{-6}$\\
\hline
\end{tabular}
\caption{Timing and noise parameters for the eSHe architecture described 
in~\cite{Chi:tailoring}.
}
\label{tab:parameters}
\end{center}
\end{table}

Our MC error model follows the example of the one used in~\cite{Steane:overhead}.
Qubit error states are again represented by Pauli operators.
However, errors are generated by a random number generator (RNG) that produces
values between 0.0 and 1.0. 
If the random value is less than the noise parameter for the error event,
then an error is produced.
We use the Mersenne Twister pseudo-RNG that is bundled 
with RngPack 1.1a~\cite{rngpack}.
This RNG has an exceptionally long period of $2^{19937}-1$.
The same set of operations and error events are modeled in both error models.
Because a single iteration of the MC model yields only a boolean value of 
whether or not that run through of the program succeeded,
the MC model must be executed many times to estimate a success probability.

We executed our analytical error model results on a 12 GB,
1.8 GHz Opteron computer.
The MC results are from a slightly faster 4 GB, 2 GHz Athlon 64 X2 computers. 
Performance between these two setups are comparable and should
yield a reasonable view of performance differences in the 
two error models.

Our simulated quantum programs focus on the error recovery procedures that dominate every program.
We adopt the [[7,1,3]] QECC for this paper.
Error recovery consists of two phases: one to detect and correct
bit-flip errors, and the other, phase-flip errors.
Each of these phases consists of three syndrome extractions using
specially prepared and verified ancilla blocks.
If a majority syndrome exists and indicates an error, correction procedures
are applied.
The [[7,1,3]] QECC is only capable of correcting general errors 
with a weight one qubit per code block.
If more than one qubit in a block has an error, then the block is considered to be uncorrectable, 
and the QC is considered to have crashed.
The analytical error model tallies up all the error state probabilities that do not crash.
Likewise, the MC simulator tallies the number of iterations that do not crash.

We do not focus on higher-level program behavior for this paper.
Logical qubits are entangled with each other through logical CNots,
and every logical operation is followed by error recovery.
The final cycle of program execution is concluded with measurement of the
program qubits.

\section{Results}
\label{sec:results}

\subsection{Threshold parameter exploration}

Our metrics of interest in evaluating error models are
simulation time and accuracy.
We will present these metrics while exploring the 
effects of the various parameters available in the analytical error model:
event branch thresholds, merge thresholds, and preservation versus lossy merges.
For this parameter exploration, we direct our error model to evaluate
a simple quantum application involving two logical qubits: 
a logical CNot is performed on these two logical qubits, 
followed by error recovery of both qubits, a second CNot and measurement.
For parameter values,
we vary the event branch threshold from $10^{-5}$ to $10^{-7}$ and
the merge threshold from $10^{-10}$ to $10^{-16}$.
Smaller treshold values result in fewer applications of the treshold and 
a more accurate result.

The error model is directed to measure the crash rate of the QC:
the probability that the program will not complete successfully due to 
errors overwhelming error correction capabilities.
All of our experiments in this parameter exploration are
simulating the same quantum program.
We select the results from using the smallest threshold values
as the baseline results for evaluating accuracy,
because they should lead to the least amount of information being thrown away.
These baseline parameters are $10^{-7}$ for the event branch threshold
and $10^{-16}$ for the merge threshold.
Both preservation and lossy merge approaches resulted in nearly the same
crash rate (within 0.004\% of each other) with these threshold values:
a crash rate of $1.55\times10^{-5}$.
Our results plot relative inaccuracy among the simulated results as 
percentage deviations from the baseline result.

\begin{figure}
\centering
\includegraphics[width=4in]{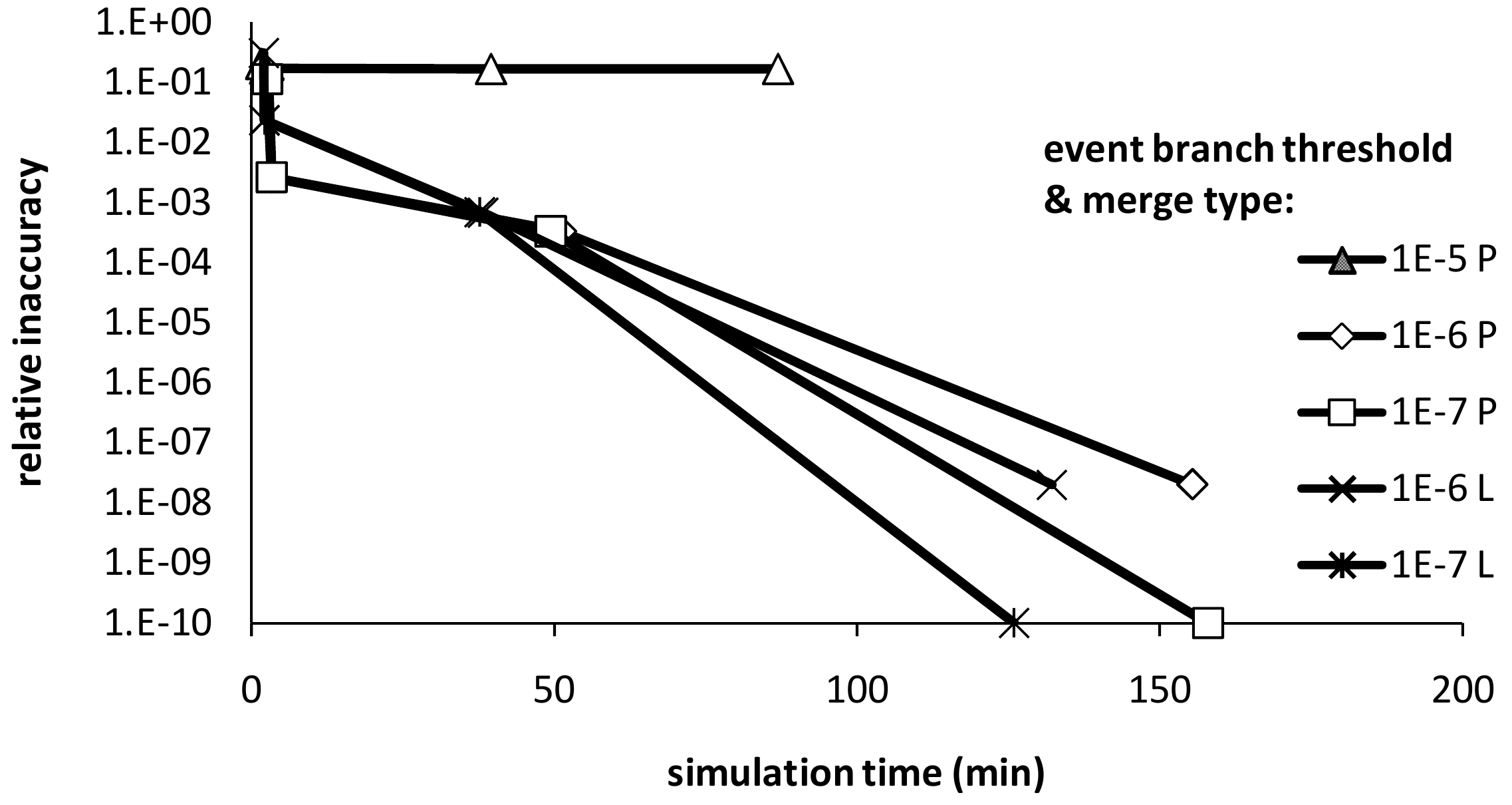}
\caption{
A parameter analysis of the analytical error model
plots inaccuracy (lower is better) as a function of simulation time.
Each line represents a particular event branch threshold
ranging from $10^{-5}$ to $10^{-7}$,
and the merge threshold varies along the line from
$10^{-10}$ to $10^{-16}$ in increments of 100
(inaccuracy declines with the merge threshold).
The legend also specifies whether the preservation (P) or lossy (L) 
approach is used for merging.
The results indicate that a low merge threshold yields the best accuracy at the cost of simulation time.
Lossy merging results in better performance than preservation merges but
results in greater inaccuracy for larger merge thresholds.
The points along the x-axis have nearly zero difference from the baseline 
results.
}
\label{fig:paramTimeAccuracy}
\end{figure}

Figure~\ref{fig:paramTimeAccuracy}
plots the relative inaccuracy versus simulation time for a variety of
threshold parameters.
Each line in the plot corresponds to a specific event branch threshold
and merge approach (preservation or lossy).
An event branch threshold of $10^{-5}$ is shown to be too large 
and results in an
inaccuracy greater than 10\% regardless of the merge threshold.
However, once the event branch threshold is $10^{-6}$ or smaller,
the merge threshold dominates both accuracy and simulation runtime.
A merge threshold of $10^{-12}$ yields a crash rate with accuracy within 0.1\%
and a simulation time of only a few minutes.
Even greater accuracy is available by reducing the merge threshold at the cost
of greatly expanding the resultant probability tree and 
increasing the simulation time to the order of 1 to 3 hours.

Lossy merges perform faster than preservation merges because
less work is performed when the merge threshold 
is exceeded.
However, the lossy merges are also significantly less accurate
than the preservation merges for larger merge threshold values
($10^{-12}$ or greater).
This accuracy difference between the two merge approaches dissipates
as the merge threshold is reduced.
By comparing the results using both merge approaches,
one may get an idea of whether further reductions in the merge threshold
may yield gains in accuracy.

\subsection{Comparison with Monte Carlo simulation}

\begin{figure}
\centering
\includegraphics[width=4in]{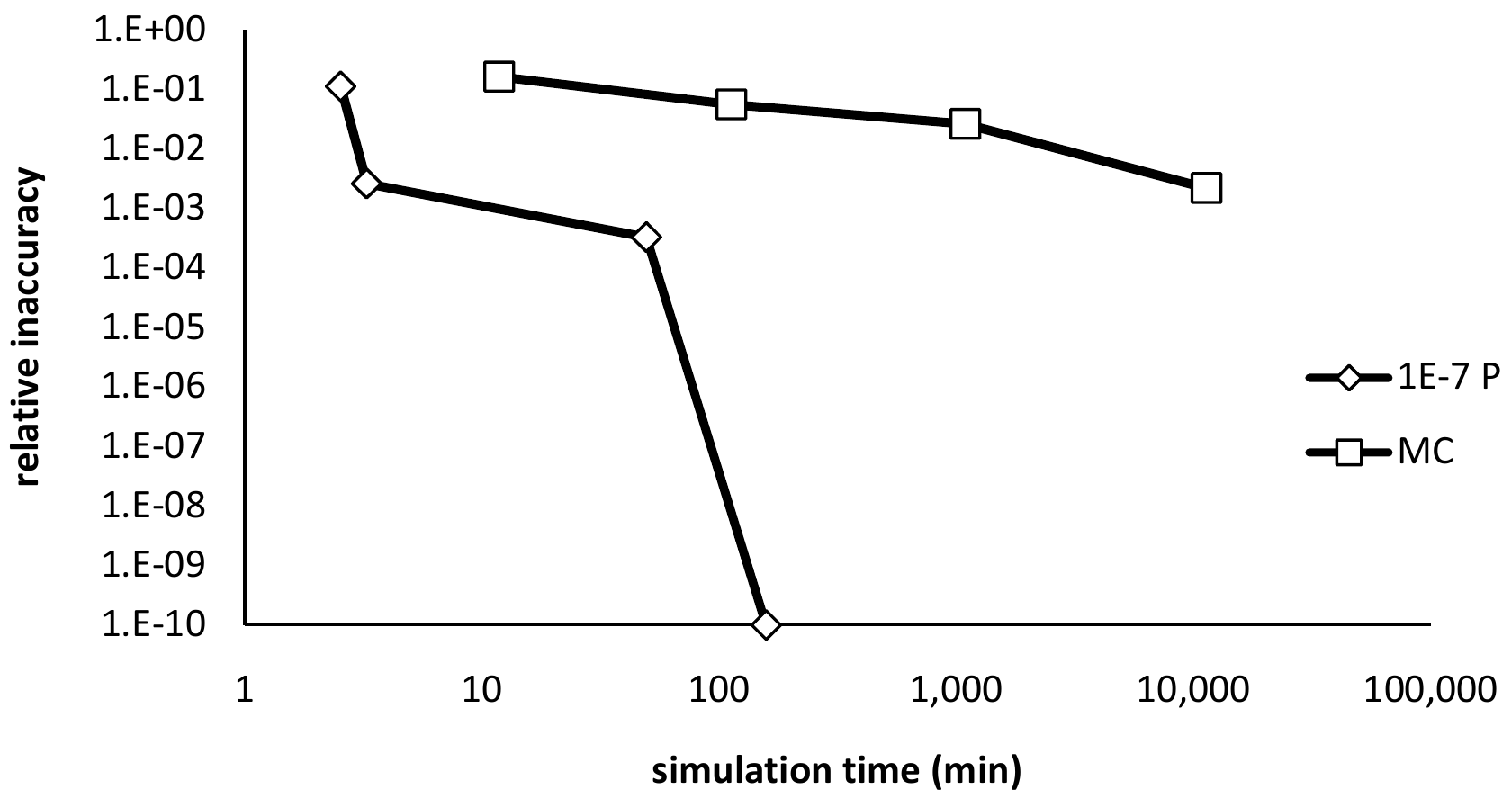}
\caption{
Comparing the analytical error model and the Monte Carlo error model
shows that the analytical model performs on the order of 1,000x better
for accuracy levels of 1\% or less.
}
\label{fig:MCcompare}
\end{figure}

Figure~\ref{fig:MCcompare} compares the analytical error model 
with the Monte Carlo error model.
These results use the same quantum program from the parameter analysis.
The analytical model results are represented by the data series
with the $10^{-6}$ event branch threshold,
using the preservation merge approach
while varying merge thresholds.
Monte Carlo results are presented with a varying number of iterations per 
data point: $10^{6}$ to $10^{9}$ iterations with interval factors of 10.
Increasing the number of iterations linearly increases the simulation time
and also increases accuracy.

The most accurate MC result is off by only 0.22\% 
(accurate to 3 significant digits) 
from the baseline.
However, this level of accuracy for the MC model comes at a great cost:
the simulation took 7.9 days to execute.
The analytical error model results with a similar level of accuracy
($10^{-7}$ event and $10^{-12}$ merge thresholds)
runs in 3.3 minutes.
This is a speedup of over 3,400X compared to the MC model 
running 1 billion iterations.

The chief advantage of the MC model is that it is simpler than the
analytical model and has
lesser development and verification costs.
Its performance scaling is also fairly straightforward and predictable:
it is linearly proportional to the number of operations and iterations.
Because it is a sampling approach, 
accuracy per iteration is proportional to the actual crash rate.
The MC model remains valuable as it may calculate
a reasonable estimate of the crash rate 
without too many iterations.
Larger MC problems may be solved by distributing the workload across 
multiple computer processors.

\subsection{Scaling with program size}

Whereas the MC model has predictable scaling properties 
with respect to program size,
the analytical error model's scaling is less certain
as it depends on interactions between the
current error map states,
the error events and tasks,
and the branch and merge thresholds.
Memory consumption is of particular concern, because
the probability tree has a worst-case exponential scaling behavior
that we attempt to mitigate with the application of branching thresholds.
The basic two-logical qubit program used earlier in the parameter exploration 
consumed about a gigabyte of memory for the analytical model 
(partially due to unhurried garbage collection in the Java virtual machine) 
compared to hundreds of megabytes for the MC simulator.
This subsection evaluates how the analytical model's 
runtime and error map sizes
scale with program size.

We extend our basic quantum program to support an arbitrary number of logical
qubits.
Given $N$ logical qubits, our scalability-testing program applies $N-1$ logical CNots and recoveries spread over $\log_2 N$ phases 
(each additional phase consumes an additional 7 cycles of program runtime).
These logical CNots are applied in a tree-like fashion with 1 CNot in 
the first phase, 2 in the second, 4 in the third, and so forth. 
The possibility of correlated errors exists among all $N$ logical qubits.
All of the logical qubits merge into the same QubitSet by the end of the program,
so this should give an indication of how the error map
scales with larger programs.

\begin{figure}
\centering
\includegraphics[width=4in]{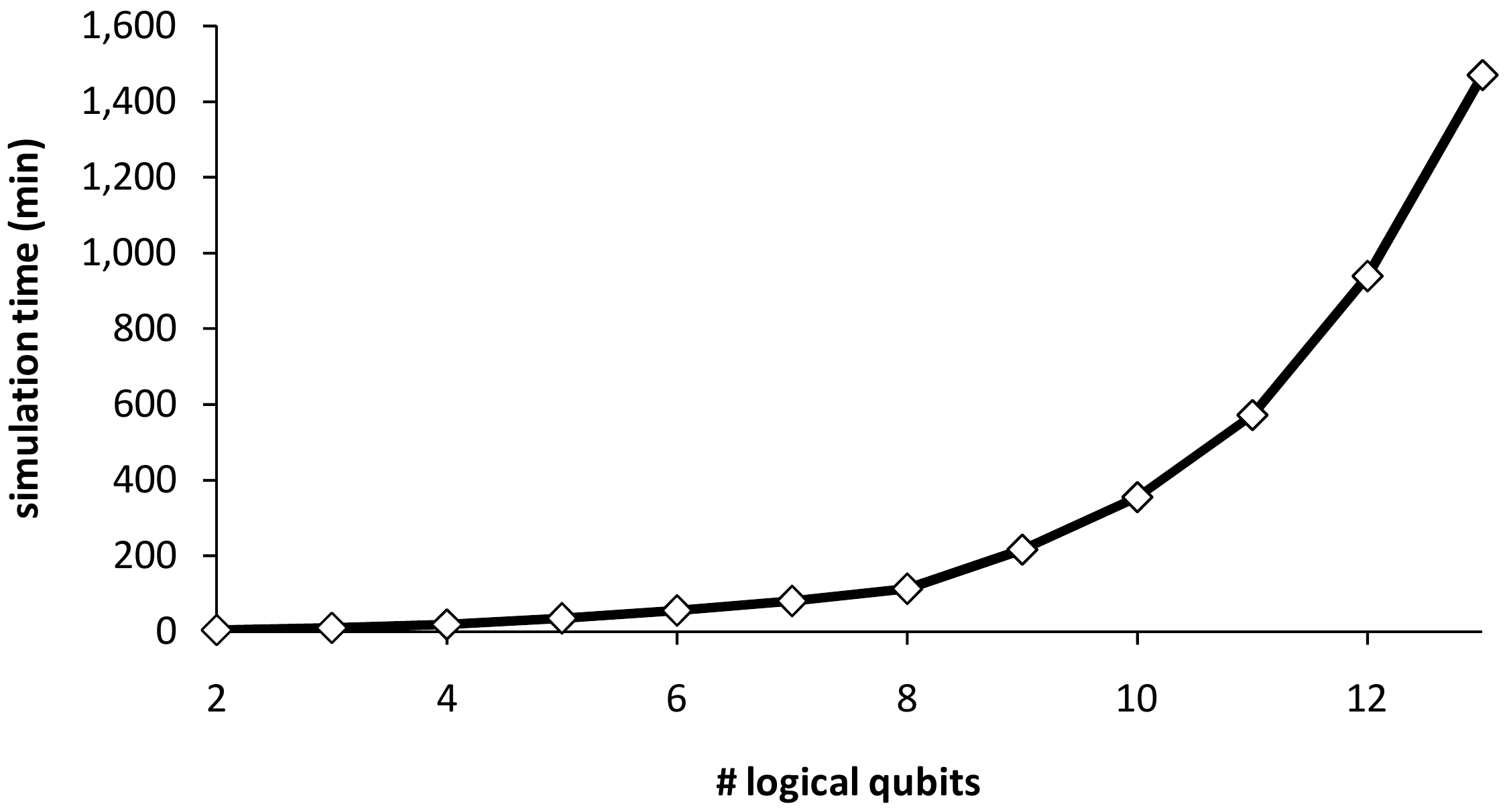}
\caption{
Simulation time for the analytical error model grows
quickly as we increase the program size (both number of logical
qubits and number of cycles).
}
\label{fig:scaleTimes}
\end{figure}

\begin{figure}
\centering
\includegraphics[width=4in]{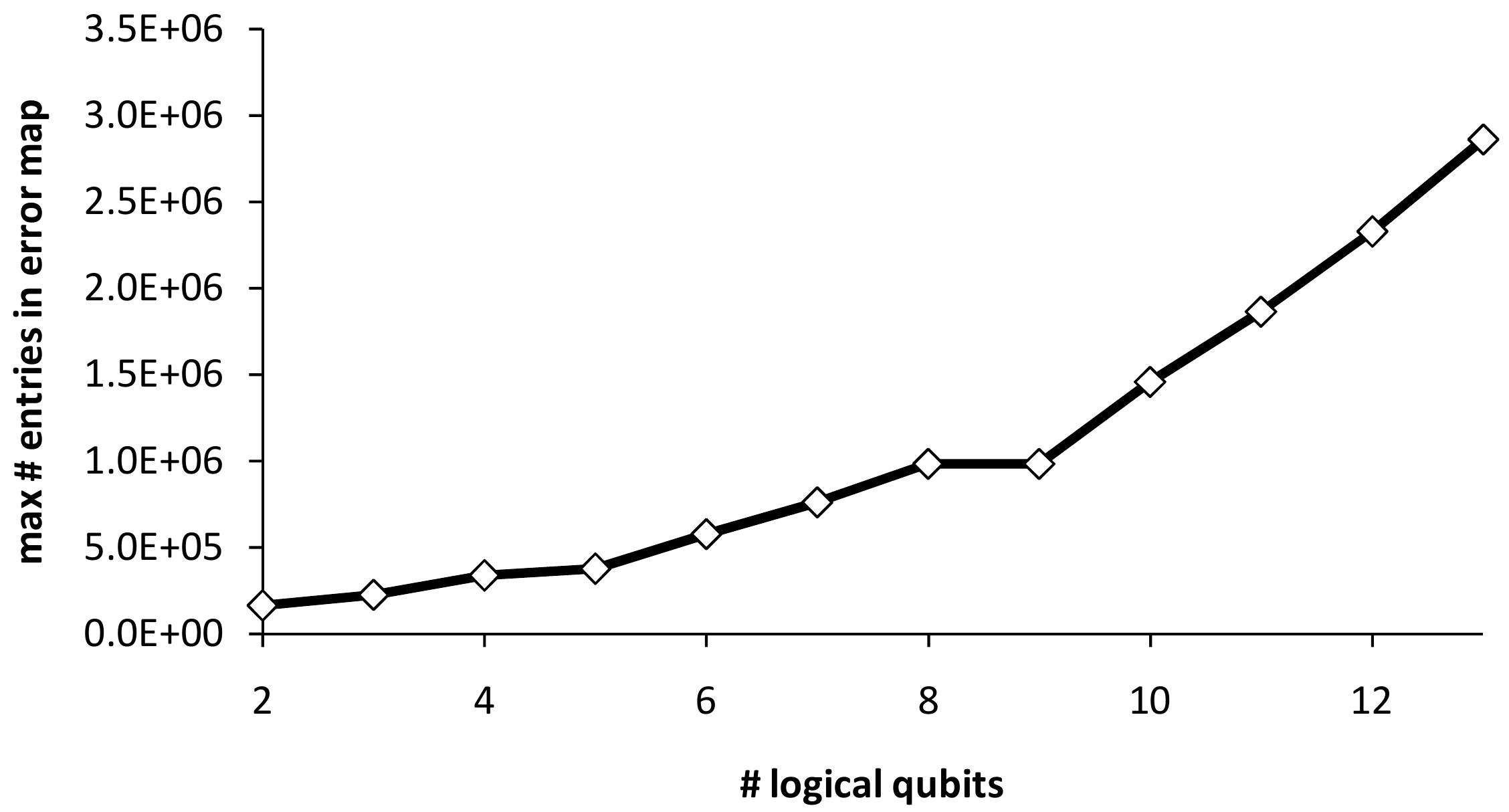}
\caption{
The maximum error map size also scales super-linearly as we scale the
number of logical qubits in the application.
}
\label{fig:scaleErrors}
\end{figure}

Figure~\ref{fig:scaleTimes} plots how the analytical error model's
runtime scales with the program size.
The event and merge threshold parameters are kept constant here at 
$10^{-6}$ and $10^{-12}$, respectively, and we utilize the preservation merging approach.
There are inflection points in this curve at 4 and 8 logical qubits;
these points correspond to when the program size is increased by an extra logical CNot phase.
Figure~\ref{fig:scaleErrors} plots how the error map size also scales 
super-linearly (albeit not quite as steeply)
with increasing program size.
This scaling suggests that runtime is based on a combination of cycle time and error map size.

The primary limiting factor to scalability is memory capacity. 
In our experiments varying the number of logical qubits from 2 through 13,
our memory consumption varied from 1 to 8~GB.
Thus, this error model is limited to studying small-scale problems on the order of a dozen logical qubits in the [[7,1,3]] QECC. 
To give some perspective on the error correction overhead, 
the scalability test with 10 logical qubits simulated a total of 1150 
physical qubits over 34 cycles.
Pairing this combinatorial noise model with a reasonably high-performance computer with 8~GB of memory would permit the efficient study of recovery procedures and simple logical circuits for the [[7,1,3]] QECC.

\subsection{Application to other QECCs}

Scalability with respect to more complicated (i.e., effective) QECCs 
is expected to be poor as this combinatorial approach scales with the number
of viable error states.
Consider replacing the [[7,1,3]] QECC with the [[21,3,5]] Golay code, which may correct errors of up to 2 qubits per 21-bit code block. 
Whereas the [[7,1,3]] code has 22 non-failing error states per code block,
the [[21,3,5]] Golay code has 1954 such states per code block.
We may estimate, then, that evaluating the Golay code will require 89x more time and memory.
With simple [[7,1,3]] experiments requiring approximately 1~GB of memory, similar [[21,3,5]] experiments would require 89~GB of memory.
Applying this model to such a problem would require either 
distributing the error map data across many computers 
or utilizing a disk-based data structure. 

\section{Conclusion}
\label{sec:conclusion}

We have presented a new analytical error model for QC simulation
that offers the potential for speedups on the order of 1,000X
compared to Monte Carlo simulation
for accurate calculations of fidelity.
Like MC simulation, our model has runtime parameters that offers the user
trade-offs between accuracy and simulation speed, 
enabling its use for a range of applications.
While the scalability of our noise model is memory-limited, 
it is capable of
analyzing interesting problems involving on the order of a dozen [[7,1,3]] logical qubits.

\bibliography{paper}

\end{document}